\documentclass{webofc}
\usepackage[varg]{txfonts}  
\usepackage{hyperref}
\usepackage{url}

\hypersetup{colorlinks=true,citecolor=blue,urlcolor=blue,linkcolor=blue}

\begin{document}

\title{Underground nuclear astrophysics: Status and recent results from
Felsenkeller laboratory}

\author{\firstname{Eliana} \lastname{Masha}\inst{1}\fnsep\thanks{\email{e.masha@hzdr.de}} \and
        \firstname{Daniel} \lastname{Bemmerer}\inst{1}\and
        \firstname{Axel} \lastname{Boeltzig}\inst{1}\and
        \firstname{Konrad} \lastname{Schmidt}\inst{1} \and
        \firstname{Anup} \lastname{Yadav}\inst{1}\and
        \firstname{Steffen} \lastname{Turkat}\inst{2} \and
        \firstname{Kai} \lastname{Zuber}\inst{2}
}

\institute{Helmholtz-Zentrum Dresden-Rossendorf, 01328 Dresden, Germany
\and Technische Universität Dresden, 01069 Dresden, Germany}

\abstract{%
For almost three decades it has been known that the study of astrophysically important nuclear reactions between stable nuclei requires the use of low-background, underground accelerator laboratories. The Felsenkeller shallow-underground laboratory in Dresden, shielded by a 45 m thick rock cover, hosts a 5 MV Pelletron ion accelerator with an external sputter ion source (mainly able to provide carbon and oxygen beams) and an internal radio-frequency ion source (providing proton and alpha beams). The reduced muon, neutron and gamma-ray background achieved both with natural and active shielding situate the laboratory well in line with deep underground accelerator labs worldwide and allows highly sensitive nuclear reaction experiments. Currently, measurements affecting the solar fusion and Big Bang nucleosynthesis are ongoing. In addition to in-house research by HZDR and TU Dresden, the lab is an open facility for scientific users worldwide, with beam time applications reviewed by an independent science advisory board. Furthermore, EU-supported transnational access is available via the ChETEC- INFRA network for nuclear astrophysics.
A brief introduction to underground nuclear astrophysics, status of the Felsenkeller shallow-underground laboratory and some preliminary results are discussed.}

\maketitle

\section{Introduction}
\label{intro}
Nuclear processes are important for the energy production and the synthesis of the elements in stars. The initial synthesis of elements, from hydrogen to beryllium, took place predominantly during the Big Bang Nucleosynthesis (BBN). However, the subsequent formation of heavier elements is a continuous process on various stages of stellar evolution and other astrophysical scenarios. Understanding these nuclear processes, and in particular measuring the cross section is the primary objective of nuclear astrophysics. This interdisciplinary field connects multiple scientific disciplines, including astronomy, atomic physics, and neutrino physics among all other fields. The nuclear processes affecting the chemical evolution of the universe are mainly dominated by charged particle-induced reactions \cite{iliadis_book}. These reactions and their role in different astrophysical scenarios depend on the temperature (energy) where these reactions occur. At astrophysically relevant energies, the cross section of these reaction is very small and is exponentially decreasing towards lower energies due to the penetrability of the Coulomb barrier. The expected reaction rate in the laboratory, considering typical direct experimental conditions, is very low (few counts per day) and usually the signal is hidden by the environmental background. Therefore, their study require low-background environments.
In this regard, the natural shielding offered by an underground laboratory provides a substantial reduction in the cosmic flux by several orders of magnitude and the experiments performed in such labs have proven great effectiveness. 
The LUNA experiment (Laboratory for Underground Nuclear Astrophysics), located at Gran Sasso National laboratory (LNGS), overburden by about 1400 m (3800 meters of water equivalent) is the pioneering experiment for underground nuclear astrophysics measurements. At LUNA the muon and
neutron fluxes are suppressed by six and three orders of magnitude, respectively, compared with the Earth’s surface \cite{Broggini_18}.
In more then 30 years, LUNA with the installation of a 50 kV and 400 kV accelerator, both providing proton and helium beams, has studied several nuclear reactions important for Big Bang nucleosynthesis (BBN), Solar hydrogen burning, the CNO and NeNa cycles, as well as slow neutron capture processes. Some recent results are reported in \cite{Mossa_20, Cavanna_15, Piatti_22, Ciani_21}. 
Very recently a new 3.5 MV accelerator \cite{MV_19} providing proton, helium and carbon beams will enable the study of nuclear reactions involved in the carbon burning phase of stars and therefore improve our understanding of the production of heavier elements.
Based on the long term success of the LUNA collaboration, other underground accelerators such as CASPAR 1 MV accelerator in the Sanford Underground Research Facility in the Homestake mine, South Dakota~\cite{caspar_16}, the JUNA 0.4 MV accelerator at Jinping Underground Facility in China~\cite{juna_16} and the Felsenkeller 5 MV shallow-underground ion accelerator in Germany~\cite{Bemmerer_24} are nowadays operative and measuring crucial nuclear reactions for astrophysics.
\section{Felsenkeller 5 MV shallow-underground accelerator}
The Felsenkeller shallow-underground laboratory is situated 5 km away from Dresden's city center, Germany. The labs are part of nine horizontal tunnels, two of which are used for nuclear astrophysics experiments. The 45 meters (140 meters water equivalent (m.w.e.)) of hornblende monzonite rock shields the laboratory from cosmic rays making it suitable for measurements of nuclear process of astrophysical interest. 
\begin{figure}[h]
\centering
\includegraphics[width=8cm,clip]{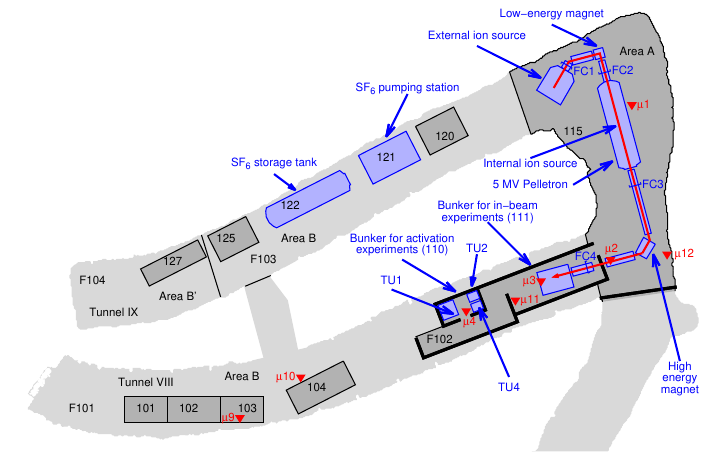}  
\caption{Layout of the underground installations in Felsenkeller tunnels VIII and IX. Area A is an enclosed radiation controlled area, the other areas are open tunnel but closed to public access. Thick black lines denotes the in-beam experimental and the activation experiment areas. The thick red line indicates the ion beam path in tandem mode. Red triangles denote the  specific areas where  muon flux in the lab has been measured.}
\label{fig-1}  
\end{figure}
Extensive examinations of muon and neutron background levels have been conducted in different areas of the laboratory before and after the opening. The muon flux is reduced by a factor of 40  with respect to the surface of Earth labs and coupled with passive shieldings and offline analysis technique is only a factor of 4-5 higher than deep underground laboratories~\cite{Jordan-canfranc_20}, \cite{Ludwig_19}. A detailed study is described in upcoming publication~\cite{Bemmerer_24}. The neutron flux in the in beam and activation experimental areas is 170-200 times lower than at the surface of the Earth~\cite{Grieger_20}. The $\gamma$-ray background, which are of particular interest for nuclear astrophysics experiments  is only 3 times higher compared to deep underground laboratories~\cite{Szucs_19}.

The 5 MV accelerator is installed in area A (Fig.~\ref{fig-1}) which connects tunnel XVIII and IX of the laboratory. The accelerator is equipped with an external sputter ion source able to produce carbon and other heavy ion beams with beam currents up to 30 $\mu$A and an internal radio frequency ion-source which produces high intensity hydrogen and helium beams. The ion beam produced from the accelerator is delivered into the in-beam experiment bunker (see Fig. \ref{fig-1}). Currently, at the end of the beamline a solid target station  with water or liquid-nitrogen cooling is adopted for most of the ongoing experiments. The target chamber is surrounded by several high-purity germanium (HPGe) detectors that can be also coupled with bismuth germanate (BGO) detectors for coincidence analyses. A full list of available detectors and setups at Felsenkeller will be reported in~\cite{Szucs_19, Bemmerer_24}.
Besides the in-beam setup the laboratory hosts one of the most sensitive HPGe detectors (163$\%$ relative efficiency \footnote{The efficiency is given relative to a 3”×3” NaI detector at 25 cm distance at 1.33 MeV gamma energy.}) for low radioactivity measurements \cite{Turkat_23}.

Felsenkeller laboratory can provide beam time by submitting a scientific project to an independent external scientific advisory board. Moreover, the facility is an open facility, part of the European ChETEC-INFRA Transnational Access (TNA)\footnote{Web page \url{www.ChETEC-INFRA.eu}, 2021-2025}. The access for beam time through ChETEC-INFRA, which might include partial travel funds, is possible every 3 months and the evaluation is done by a dedicated User Selection Panel of the ChETEC-INFRA project.

\section{Ongoing and planned experiments}\label{sec-2}
\subsection{$\mathrm{^3He(\alpha,\gamma)^7Be}$ reaction}\label{sec-2a}
The $\mathrm{^3He(\alpha,\gamma)^7Be}$ reaction is mainly relevant for two different scenarios: it can affect the $^7$Li production during Big Bang nucleosynthesis, and it controls the branching between the pp-1 and pp-2 chains in solar hydrogen burning. The BBN energy range (160 - 380 keV) has been directly explored by the LUNA collaboration \cite{Bemmerer_06} reducing the overall systematic uncertainties.
At solar energies (19 - 30 keV), the reaction cross sections drops significantly making unfeasible any direct measurement. The only way to access the solar Gamow peak is through indirect methods or theoretical extrapolations which rely on high energy data reported in \cite{Bordeanu_13} and the references therein. Recent theoretical extrapolation \cite{Neff} shows that there is a correlation between the $\gamma$-ray angular distribution of  $^3$He($\alpha,\gamma$)$^7$Be and the S(0)-factor extrapolations.
Because of the uncertainty in these extrapolations, the determined solar reaction rate has an error of 5.4$\%$. Therefore, to constrain the role of this reaction in the solar model and compare with direct observations which have reached 1$\%$ uncertainties $\gamma$-ray angular distribution measurements have been carried out at Felsenkeller laboratory. The experiment was performed using alpha beam (in the energy range $E_\textrm{CM}$\,=\,450\,-\,1220\,keV) impinging on solid $\mathrm{^{3}He}$ implanted target and 20 HPGe detectors at angles between $\theta$\,=\,25\,-\,140\,$^\circ$ with respect to the beam direction. Preliminary results show that the backward emission is systematically preferred with respect to the theory. The data analysis is ongoing and more details on the step and the analysis are reported in \cite{turkat_thesis}.

\subsection{ $\mathrm{^2H(p,\gamma)^3He}$ reaction}\label{sec-2b}
The $^2$H($p,\gamma$)$^3$He reaction is one of the main processes responsible for the deuterium destruction during the BBN and affects the primordial deuterium abundance ratio (D/H).

The primordial abundances can be directly obtained by direct observations which have reached very high precision \cite{Cooke_18} or can be computed using the BBN theory. This last one, depends on the cross-section of nuclear reactions involved and has only one single free parameter, the baryon-to-photon ratio,which can also be extracted from the CMB measurements \cite{plank_18}.
For several years the uncertainties in the primordial deuterium abundances were primarily dominated by the  $\mathrm{^2H(p,\gamma)^3He}$ reaction. However, this reaction was directly measured within BBN energy range by the LUNA collaboration \cite{Mossa_20}, using a windowless gas target. The LUNA results reduced the cross section uncertainties down to 3$\%$ and provided constraints for the cosmological parameters comparing the precise primordial deuterium abundance from the BBN model with direct observations \cite{Mossa_20}. Nevertheless, the new LUNA S-factor extrapolations are not fully in agreement with theoretical {\it ab initio} calculations\cite{Marcucci}. Recently, the $^2$H($p,\gamma$)$^3$He reaction was measured at high energies \cite{turkat_21} using implanted deuterium targets on tantalum backing, revealing a different high-energy extrapolation of the S-factor compared to the previous LUNA data\cite{Mossa_20}.

To address these discrepancies and to constrain theoretical {\it ab initio} calculations, a new experimental campaign was devoted last year at Felsenkeller laboratory using solid deuterium (ZrD2) targets and several HPGe detectors at seven different angles. During this campaign the $\mathrm{^2H(p,\gamma)^3He}$ reaction emits single $\gamma$-rays with an energy between 5.5 and 6.8 MeV. In this region of interest the experiment fully exploits the cosmic background suppression at Felsenkeller lab. The $\gamma$-ray detection efficiency is carefully evaluated using standard radioactive sources, $^{60}$Co, $^{137}$Cs, $^{88}$Y and exploiting the well known $\mathrm{^{27}Al(p,\gamma)^{28}Si}$ resonance, at $E$\textsubscript{p}$\,=\,992\,$keV.

This setup allows the measurement of absolute cross section and the $\gamma$-ray angular distribution at different angles in the proton beam energy range 300-1200 keV. This experimental campaign will effectively bridge the energy range gap between existing low and high-energy literature data. 
\begin{figure}[h]
\centering
\includegraphics[width=10cm,clip]{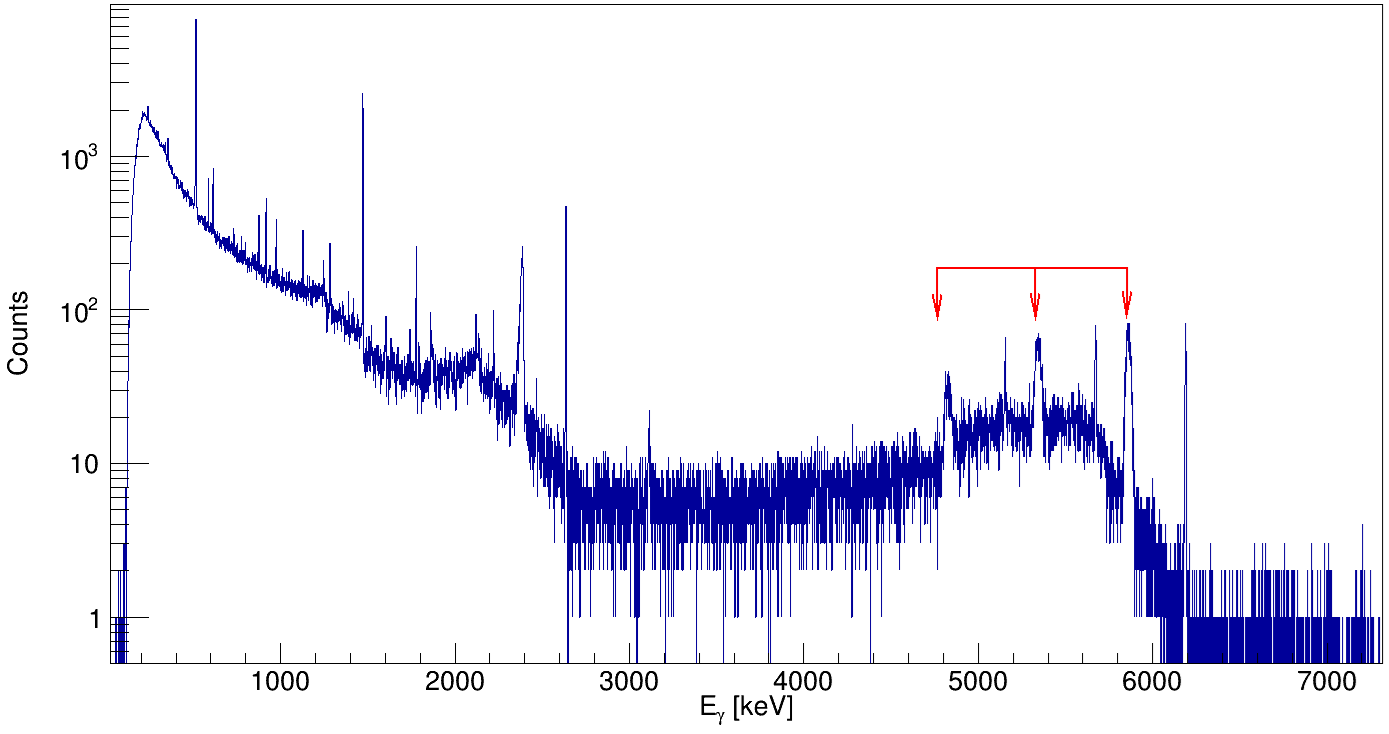}  
\caption{$\mathrm{^2H(p,\gamma)^3He}$ $\gamma$-ray spectrum measured at 450 keV beam energy. The full energy peak together with first and second escape peaks are given by the red arrows. }
\label{fig-2}      
\end{figure}
Fig. \ref{fig-2} shows a typical gamma ray spectrum taken with the 90$^\circ$ detector at proton beam energy of 450 keV. To have under control the target stability, beside the ERDA analysis of the targets before and after the experimental campaign, for each measured energy, reference runs at proton energy of 600 keV were performed. The analysis and comparison of the $\gamma$-ray angular distribution with theoretical calculations are ongoing.

\subsection{$\mathrm{^{12}C(p,\gamma)^{13}N}$ reaction}\label{sec-2c}

The $\mathrm{^{12}C(p,\gamma)^{13}N}$ reaction is part in the CNO cycle of hydrogen burning. The $\mathrm{^{12}C(p,\gamma)^{13}N}$ reaction has an important role in determining the abundance of the stable carbon $\mathrm{^{12}C}$ isotope, fundamental element in our life. In stars, in asymptotic giant and giant branches (AGB and RGB stars) the H-shell $^{12}$C/$^{13}$C isotopic ratio is commonly used to trace stellar nucleosynthesis and the galactic evolution\cite{Savage_01}. 
Due to poorly constrained extrapolations at the relevant energies, the astrophysical models were unable to reproduce the observed $^{12}$C/$^{13}$C isotope ratio. A recent measurement, directly at the Gamow energies of interest,  was performed at LUNA at energies below 400 keV using a high-efficiency BGO detector, which is optically segmented into six individual crystals, each covering an angle of 60 degrees, allowing a 4$\pi$ configuration geometry \cite{Skow_luna23}.

\begin{figure}[h]
\centering
\includegraphics[width=10cm,clip]{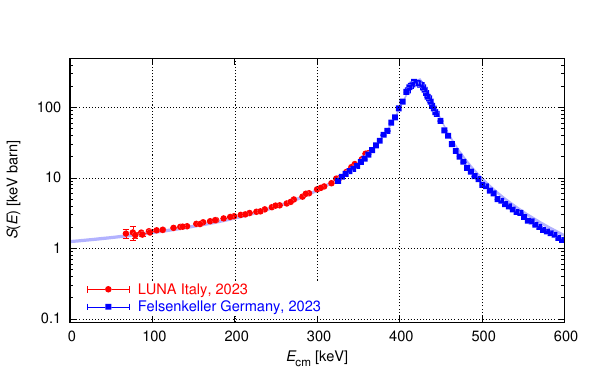}  
\caption{$\mathrm{^{12}C(p,\gamma)^{13}N}$ S-factors. In blue the data measured at LUNA \cite{Skow_luna23}, in red the results from the Felsenkeller laboratory \cite{Skow_fels23}. The light blue line is the R-matrix extrapolation computed using all the available literature data for the $\mathrm{^{12}C(p,\gamma)^{13}N}$ reaction.}
\label{fig-3}       
\end{figure}

At Felsenkeller the measurement was performed at higher energies,  $E_\textrm{p}$= 350-670 keV using evaporated carbon targets on tantalum discs. The targets were liquid-nitrogen cooled and the degradation was continuously monitored using the peak-shape analysis and the Nuclear Resonant Reaction Analysis.
The $\gamma$-ray angular distribution  in the region of interest reported here was expected to be isotropic \cite{rolfs74}. However, thanks to the setup available at Felsenkeller, this component was also checked at $E_\textrm{p}$= 400, 464 and 555 keV. More details on the data taking and analysis are reported in Ref. \cite{Skow_fels23}. 
The high energy data measured at Felsenkeller \cite{Skow_fels23} are fully in agreement with the low energy data measured at LUNA \cite{Skow_luna23} (Fig. \ref{fig-3}).
Considering these data, together with all available literature data, S factor extrapolations down to energies of astrophysical interest were obtained adopting the R-matrix analysis (AZURE2 code \cite{azure}). Results are shown in Fig. \ref{fig-3}.

\subsection{Future developments and experiments}\label{sec-2d}
Within a comparatively short time Felsenkeller laboratory is contributing significantly to the nuclear astrophysics community. At the time of this writing, stability tests of solid carbon targets under intensive carbon ion beam are ongoing. These targets will be used by the LUNA collaboration for the study of the $^{12}$C +$^{12}$C fusion reactions which play an important role in explosive carbon burning energies of astrophysical interest. 
The experimental plan at the Felsenkeller 5MV accelerator involve a wide selection of nuclear reactions. These include the study of the $\mathrm{^{13}C(p,\gamma)^{14}N}$ reaction, the investigation of low-energy resonances in the $\mathrm{^{14}N(\alpha,\gamma)^{18}F}$ reaction, and exploration of resonances and direct capture mechanisms in the $\mathrm{^{15}N(\alpha,\gamma)^{19}F}$. Additionally, one of the main experiments is the  study of the $\mathrm{^{12}C(\alpha,\gamma)^{16}O}$ reaction. This reaction plays a crucial role in stellar helium burning processes and the synthesis of heavier elements and currently there are no data below 1 MeV center-of-mass energy, relevant for astrophysics. To study this reaction, a new \(^{4}\text{He}\) gas-jet target is under development at Felsenkeller \cite{Yadav_23}.
The data collection phase for certain experiments has been completed, with analysis ongoing, while others are scheduled for future beam time allocation.


\begin{thebibliography}{}

\bibitem{iliadis_book}
C. Iliadis, Nuclear Physics of Stars, 2nd ed. (Wiley-VCH, Weinheim, 2015)
\bibitem{Broggini_18}
C. Broggini, et al., Prog. Part. Nucl. Phys. 98, 55 (2018).
\bibitem{Mossa_20}
V. Mossa et al., Nature 587, 210 (2020).
\bibitem{Cavanna_15}
F. Cavanna, et al., Phys. Rev. Lett. 115 252501 (2015)
\bibitem{Ciani_21}
G.F. Ciani et al., Phys. Rev. Lett. 127, 152701 (2021)
\bibitem{Piatti_22}
D. Piatti et al., Eur. Phys. J. A  58(10), 194 (2022)
\bibitem{MV_19}
A. Sen et al., Nucl. Inst. Meth. B 450, 390 (2019).
\bibitem{caspar_16}
D. Robertson, et al., Eur. Phys. Journal Web of 935 Conf. Vol. 109, p. 09002, (2016).
\bibitem{juna_16}
W. Liu et al., Science China Physics, Mechanics, and Astronomy 59, 5785 (2016).
\bibitem{Bemmerer_24}
D. Bemmerer et al., submitted to Eur. Phys. J. A (2024) 
\bibitem{Jordan-canfranc_20}
D. Jordan et al., Astropart. Phys. 118, 102372 (2020).
\bibitem{Ludwig_19}
F. Ludwig et al., Astropart. Phys. 112, 24 (2019).
\bibitem{Grieger_20}
M. Grieger et al., Phys. Rev. D 101, 123027 (2020).
\bibitem{Szucs_19}
T. Szücs et al., Eur. Phys. J. A 55, 174 (2019)
\bibitem{Turkat_23}
S. Turkat et al., Astropart. Phys. 148, 102816 (2023)
\bibitem{Bemmerer_06}
D. Bemmerer, et al., Phys. Rev. Lett. 97 (2006)
\bibitem{Bordeanu_13}
C. Bordeanu, et al., Nucl. Phys. A 908, 1 (2013)
\bibitem{Neff}
T. Neff, Phys. Rev. Lett. 106 (2011)
\bibitem{turkat_thesis}
S. Turkat, PhD thesis, Technische Universität Dresden, 2023.
\bibitem{Cooke_18}
R. J. Cooke, et al., The Astrophysical Journal 830, 148 (2016).
\bibitem{plank_18}
P. A. R. Ade, et al., arXiv:1502.01589v2.
\bibitem{Marcucci}
L. Marcucci et al., Phys. Rev. Lett., 116 (2016).
\bibitem{turkat_21}
S. Turkat et al., Phys. Rev. C 103, 045805 (2021)
\bibitem{Savage_01}
C. Savage,et al.,  American Astronomical Society Meeting Abs., p. 59.13, May 2001.
 \bibitem{Skow_luna23}
J. Skowronski et al., Phys. Rev. Lett. 131, 162701945 (2023)
\bibitem{rolfs74}
 C. Rolfs and R. E. Azuma, Nucl. Phys. A 227, 291-308 (1974)
\bibitem{Skow_fels23}
 J. Skowronski et al., Phys. Rev. C 107, L062801 (2023).
\bibitem{azure}
R. Azuma et al., Phys. Rev. C 81, 045805 (2010).
\bibitem{Yadav_23}
A. Yadav et al., EPJ Web of Conferences 279, 13002 (2023)

\end{thebibliography}
\end{document}